# Calculation of Confined Phonon Spectrum in Narrow Silicon Nanowires using the Valence Force Field Method

Hossein Karamitaheri[1,2], Neophytos Neophytou[1], Mohsen Karami Taheri[3], Rahim Faez[2], and Hans Kosina[1]

[1]Institute for Microelectronics, TU Wien, Gußhausstraße 27-29/E360, A-1040 Wien, Austria

[2]School of Electrical Engineering, Sharif University of Technology, Tehran, Iran

[3] Department of Computer Engineering, Naragh Branch, Islamic Azad Universitry, Naragh, Iran

E-mail: {karami | neophytou | kosina}@iue.tuwien.ac.at

## Abstract

We study the effect of confinement on the phonon properties of ultra-narrow silicon nanowires of side sizes of $1-10\text{nm}$. We use the modified valence force field method to compute the phononic dispersion, and extract the density of states, the transmission function, the sound velocity, the ballistic thermal conductance and boundary scattering-limited diffusive thermal conductivity. We find that the phononic dispersion and the ballistic thermal conductance are functions of the geometrical features of the structures, i.e. the transport orientation and confinement dimension. The phonon group velocity and thermal conductance can vary by a factor of two depending on the geometrical features of the channel. The <110> nanowire has the highest group velocity and thermal conductance, whereas the <111> the lowest. The <111> channel is thus the most suitable orientation for thermoelectric devices based on Si nanowires since it also has a large power factor. Our findings could be useful in the thermal transport design of silicon-based devices for thermoelectric and thermal management applications.

**Index terms:** Confined phonons, silicon nanowires, lattice thermal conductance, modified valence force field method, Landauer formula.

# I. Introduction

Low dimensional silicon nanowires (NWs), and silicon-based ultra-thin layers have attracted significant attention as efficient thermoelectric (TE) materials after it was demonstrated that the thermal conductivity in such materials can be drastically reduced compared to bulk Si. Values as low as $\kappa = 2\text{W/mK}$ (compared to $\kappa = 140\text{W/mK}$ for bulk Si) were achieved [1, 2]. This resulted in $ZT$ values close to $ZT \sim 0.5$, a large improvement compared to $ZT$ of bulk Si, $ZT_{\text{bulk}} \sim 0.01$ [1, 2]. The large reduction in the thermal conductivity was attributed to enhanced scattering of phonons on the surfaces of the nanochannels.

Numerous studies can be found in the literature regarding the thermal conductivity of Si NWs [3, 4, 5, 6, 7]. The effects of different scattering mechanisms, i.e. surface roughness scattering, mass doping, phonon-phonon scattering, and phonon-electron scattering have been investigated by several authors [8, 9, 10, 11]. In these works, it is demonstrated that the thermal conductivity in ultra-narrow Si NWs drastically degrades once the diameter of the NW is reduced below 50nm, or when scattering centers are incorporated. For even smaller NW diameters, i.e. below 10nm, the effect of confinement could further change the phonon spectrum significantly, and provide an additional mechanism in the reduction of the thermal conductivity [12]. This could provide additional benefits to the thermoelectric figure of merit $ZT$. In this work, we employ the modified valence force field (MVFF) method [13] to address the effect of structural confinement on the phonon dispersion, group velocity and ballistic thermal conductance of ultra-narrow Si NWs of diameters below 10nm. We consider different transport orientations, and different cross sectional sizes. We find that although the density of phonon modes does not show any orientation dependence, the phonon group velocities and sound velocities are strongly anisotropic. This results in a higher thermal transmission in the case of the <110> NW, which results in a higher ballistic thermal conductance compared to differently oriented NWs as well. On the other hand, the <111> oriented NW has the lowest thermal conductance, almost 2x lower than that of the <110> NW, and could be the optimal choice for NW TE devices. Our results, therefore, indicate that the proper choice of the NW geometry can provide a further reduction in the thermal conductivity of NW TE devices and consequently higher *ZT* values.

## II. Approach

For the calculation of the phononic bandstructure we employ the modified valence force filed method [13]. In this method the interatomic potential is modeled by the following bond deformations: bond stretching, bond bending, cross bond stretching, cross bond bending stretching, and coplanar bond bending interactions [13]. The model accurately captures the bulk Si phonon spectrum as well as the effects of confinement [14].

In the MVFF method, the total potential energy of the system is defined as [14]:

$$U \approx \frac{1}{2} \sum_{i \in N_A} \left[ \sum_{j \in nn_i} U_{bs}^{ij} + \sum_{j,k \in nn_i}^{j \neq k} \left( U_{bb}^{jik} + U_{bs\text{-}bs}^{jik} + U_{bs\text{-}bb}^{jik} \right) + \sum_{j,k,l \in COP_i}^{j \neq k \neq l} U_{bb\text{-}bb}^{jikl} \right]$$ (1)

where $N_A$, $nn_i$, and $COP_i$ are the number of atoms in the system, the number of the nearest neighbors of a specific atom *'i'*, and the coplanar atom groups for atom *'i'*, respectively. $U_{bs}$, $U_{bb}$, $U_{bs\text{-}bs}$, $U_{bs\text{-}bb}$, and $U_{bb\text{-}bb}$ are the bond stretching, bond bending, cross bond stretching, cross bond bending stretching, and coplanar bond bending interactions, respectively [14].

In this formalism we assume that the total potential energy is zero when all the atoms are located in their equilibrium position. Under the harmonic approximation, the motion of atoms can be described by a dynamic matrix as [15]:

$$D = [D_{3 \times 3}^{ij}] = \left[ \frac{1}{\sqrt{M_i M_j}} \begin{cases} D_{ij} & i \neq j \\ -\sum_{l \neq i} D_{il} & i = j \end{cases} \right]$$

(2)

where dynamic matrix component between atoms *'i'* and *'j'* is given as shown in Ref. [14] by:

$$D_{ij} = \begin{bmatrix} D_{xx}^{ij} & D_{xy}^{ij} & D_{xz}^{ij} \\ D_{yx}^{ij} & D_{yy}^{ij} & D_{yz}^{ij} \\ D_{zx}^{ij} & D_{zy}^{ij} & D_{zz}^{ij} \end{bmatrix}$$

(3)

and

$$D_{mn}^{ij} = \frac{\partial^2 U_{\text{elastic}}}{\partial r_m^i \partial r_n^j}, \quad i, j \in N_A \text{ and } m, n \in [x, y, z]$$

(4)

is the second derivative of the potential energy after atoms 'i' and 'j' are slightly displaced along the m-axis and the n-axis ($\partial r_m^i$ and $\partial r_n^j$), respectively. To calculate $U_{\text{elastic}}$, we added only those terms of Eq. 1 that include atoms 'i' and 'j'. All other terms cancel each other out when we numerically calculate the second derivative of potential energy.

After setting up the dynamic matrix, the following eigenvalue problem is solved to calculate the phononic dispersion:

$$D + \sum_l D_l \exp(i\vec{q}.\Delta\vec{R_l}) - \omega^2(q)I = 0$$

(5)

where $D_l$ is the dynamic matrix representing the interaction between the unit cell and its neighboring unit cells separated by $\Delta\vec{R_l}$ [15]. Using the phononic dispersion, the density of states (DOS) and the ballistic transmission (number of modes at given energy) are calculated as shown in Ref. [16] by:

$$DOS(\omega) = \sum_i DOS_i(\omega) = \sum_i \sum_q \delta(\omega - \omega_i(q))$$

(6)

and

$$\overline{T}_{\text{ph}}(\omega) = M(\omega) = \frac{h}{2} \sum_{i,q} \delta(\omega - \omega_i(q)) \frac{\partial \omega_i(q)}{\partial q}$$

(7)

Once the transmission is obtained, the ballistic lattice thermal conductance is calculated within the framework of the Landauer theory as [17]:

$$K_l = \frac{1}{h} \int_0^\infty \overline{T}_{\text{ph}}(\omega) \hbar\omega \left(\frac{\partial n(\omega)}{\partial T}\right) d(\hbar\omega)$$

(8)

It is well known that the thermal conductivity in Si nanowires of very narrow cross sections is dominated by boundary scattering, which is the reason of their improved thermoelectric performance as well [2, 11, 18, 19]. The diffusive thermal conductivity can be calculated using the phonon lifetime approximation in the phononic Boltzmann transport equations as [18]:

$$\kappa_l = k_B \sum_{i,q} \tau_i(q) v_{g,i}(q)^2 \left[ \frac{\hbar \omega_i(q)}{k_B T} \right]^2 \frac{e^{\hbar \omega_i(q)/k_B T}}{\left( e^{\hbar \omega_i(q)/k_B T} - 1 \right)^2} \qquad (9)$$

where $v_{g,i}(q)$ is the group velocity of phonons of wavevector $q$ in subband $i$, given by $v_{g,i}(q) = \partial \omega_i(q)/\partial q$, and $\tau_i(q)$ is the scattering time. Here we compute the boundary-scattering-limited thermal conductivity by evaluating the boundary scattering time as:

$$\frac{1}{\tau_{B,i}(q)} = \left( \frac{1-P}{1+P} \right) \frac{v_{g,i}(q)}{d} \qquad (10)$$

where $d$ is the effective diameter (defined as $d = \sqrt{2}W$ where $W$ is the side length of the NW cross section) of the NWs and $P$ is the specularity parameter that takes values form 0 to 1, determined by the details of the surface and surface roughness [18, 20]. For a smooth surface $P = 1$, the scattering time is infinite (meaning zero scattering rates on the surface), and boundary scattering is fully specular. On the other hand, for a very rough surface $P = 0$, which results in fully diffusive boundary scattering. Recently, studies showed that experimental results for boundary scattering in narrow NWs are only explained once an almost fully diffusive boundary is assumed [21, 22].

## III. Results and Discussion

In this work we examine the effects of transport orientation and confinement length scale on the thermal properties of Si-NWs with square cross sections ($W = H$) and side sizes up to $10 \text{nm}$. We study the phononic dispersion, density of states, ballistic transmission, average group velocity, and lattice thermal conductivity of different structures.

The phononic dispersions of Si NWs of $2\text{nm} \times 2\text{nm}$ cross section in different transport orientations, <100>, <110>, and <111>, are shown in Fig. 1. There are differences in the dispersions, especially in the low energy, low momentum region, which indicates that the thermal properties could be orientation dependent as well. Figure 2 shows the ballistic transmission function of a different set of NWs, of side sizes $W = H = 6\text{nm}$ in the same three orientations, calculated as indicated above in Eq. 7. The transmission function of the <110> NWs is the highest in most part of the energy spectrum, whereas the transmission of the <111> NWs is the lowest in almost the entire energy spectrum. As a result, the ballistic lattice thermal conductance of the NWs (see Fig. 3-a) calculated using the Landauer formula as specific in Eq. 8, shows that the <110> NWs has the highest thermal conductance compared to the NWs oriented along the <100> and <111> transport directions. The thermal conductance is the lowest for the <111> NWs of all the side sizes we have considered, all the way up to 10nm as shown in Fig. 3-a. The difference between the thermal conductance of the <110> NW which has the highest performance, and the <111> NW which has the lower performance, is a factor of ~2. Another observation is that the thermal conductance increases as the cross section of the NW increases. This is expected since the larger NWs contain more transport modes. The increase is close to linear. Once the conductances are normalized by the cross section area of the NWs, however, the resultant normalized conductances remain almost constant. This is indicated in the inset of Fig. 3-a. In this case, again, the <111> has clearly the lowest conductivity, almost 2 times lower than the <110> NW for all cross section sizes. In Fig. 3-b we show the boundary-scattering-limited lattice thermal conductivity for the nanowires of 2nm and 10nm cross section sides, versus the specularity parameter *P*. As *P* increases from 0 (fully diffusive surfaces) to 0.5 (semi-diffusive surfaces), the conductivity naturally increases. The scattering is stronger for narrower NWs. For the NWs with cross section sizes of 2nm, thermal conductivities as low as ~2 W/mK are achieved for *P*=0, raising to ~5 W/mK for *P*=0.5. For NWs with cross section sizes of 10nm, the thermal conductivity is almost 5 times higher, at most ~30 W/mK. These values are significantly lower than the thermal conductivity of bulk Si ($140\text{W/mK}$), and once other scattering mechanisms are considered (i.e. phonon-phonon scattering and impurity scattering) they could lower even further. Interestingly, the order of the conductivities for NWs of different orientations is the same as in the ballistic case of Fig. 3-a, with the <110> NW having the highest and the <111> NW the lowest conductivity. The same anisotropy behavior is observed when considering only phonon-phonon scattering mechanisms [23], and different fully diffusive transport formalisms [18, 19], further demonstrating the importance of the phonon bandstructure in determining the thermal conductivity of Si NWs. For thermoelectric applications, therefore, where low thermal conductance is advantageous, the <111> NW would be the optimal choice as we show below.

We have recently studied the role of transport orientations and diameter on the thermoelectric power factor of NWs using atomistic bandstructure simulations for electron transport [24]. With regards to anisotropy, in the case of *n*-type NWs only a small orientation dependence was observed. In the case of *p*-type Si NWs, however, we showed that the <111> NWs have significantly higher power factors compared to differently orientated NWs. The <111> Si NW channel is, therefore, the most advantageous for *p*-type channel thermoelectric applications since it provides simultaneously the highest power factor and lowest thermal conductivity compared to *p*-type NWs of different transport orientations. Using the thermoelectric power factor of Ref. [24] (Fig. 3c and Fig. 7c of Ref. [24]) and the calculated thermal conductivity, in Fig. 4a and 4b we show the *ZT* figure of merit for *n*-type and *p*-type NWs, respectively, versus the NW cross section size. For *n*-type NWs, the <111> direction is the most advantageous, providing *ZT* of ~0.5. On the other hand, a *ZT* value around 1 is calculated for a very narrow *p*-type <111> NWs as shown in Fig. 4b, almost 3 times higher than that of the <110> NW. This is partly due to its higher power factor originating from its large electronic mobility [25] and partly due to its lower lattice thermal conductivity.

For the rest of the paper, we revisit the ballistic results, and attempt to elucidate the reasons behind the anisotropy of the thermal conductance and conductivity in terms of the NWs' phonon bandstructure. The ballistic transmission function is directly related to the product of the density of phonon states and phonon group velocity. Below we investigate each of these quantities and their geometry dependence in order to provide insight into the orientation dependence of the thermal conductance. For this, we consider NWs with $W = H = 6\text{nm}$ in the three different orientations under investigation. Figure 5 shows the phonon DOS, which is nearly the same for all orientations in the entire range of frequencies. This can be understood by considering the fact that the $DOS$ is mostly related to the density of atoms in the cross section of the NW. The density of atoms is mostly a bulk property and does not change drastically in Si-based structures. Each atom has three degrees of freedom ($3 \times 3$ elements in its Dynamic matrix component), and therefore provides three eigenmodes. Although for ultra-narrow structures i.e. of $H=W=2$nm there could be some differences in the DOS due to finite size effects a $6\text{nm} \times 6\text{nm}$ structure is large enough such that the DOS is orientation independent.

On the other hand, as shown in Fig. 6, the sound velocity, defined by the slope of the acoustic modes near the Brillouin zone center, is not an isotropic quantity. We note here that the 2 lowest modes in phonon dispersion of NWs are flexural modes. The third mode is the one used to calculate the transverse sound velocity, whereas the fourth mode gives the longitudinal sound velocity. Our results, as shown in Fig. 6, are in good agreement with previous studies on NW sound velocities [14, 26]. The <111> NW has the highest longitudinal sound velocity (green-solid), whereas the <100> NW the lowest sound velocity for all NW cross section sizes. The transverse velocities are lower, and in that case, the <100> NW has the highest velocity. The values calculated here are different, and in general lower than the bulk sound velocities (dots in the right side of Fig. 6). The velocities in NWs are lower, also in agreement with other reports, since confinement flattens the phonon spectrum and reduces sound velocities [27].

At first, following a bulk way of analysis, it seems that according to the ordering in the sound velocity one would expect that the thermal conductance should be higher in the <111> NWs, followed by the <110> NWs and then by the <100> NWs. As we show in Fig. 3, however, the order is different, namely that the <110> NWs have the highest thermal conductivity, followed by the <100>, and then by the <111> NWs. The fact is that the phonon dispersion is modified in nanostructures so that the bulk definition of acoustic and optical modes is no longer the relevant quantity. The sound velocity determines the dispersion only in a small part of energy spectrum ( ~ 5meV of out of the total ~ 65meV). In addition, at room temperature, all the phonons in the entire energy spectrum contribute to the thermal conduction. Therefore, the sound velocity alone cannot describe the NW phonon spectrum, which is an indication that the bulk treatment is inefficient for understanding thermal transport in NWs. What needs to be taken into consideration in order to correctly interpret the results, therefore, is the group velocity of all phonon modes in all energies. For this, we define the effective/average group velocity as follows:

$$\left\langle\left\langle v_{\mathrm{g}}(\omega)\right\rangle\right\rangle = \frac{\sum_{i,q} v_{\mathrm{g},i}(q)\delta(\omega-\omega_i(q))}{\sum_{i,q}\delta(\omega-\omega_i(q))} \quad (11)$$

where the sum holds over all the modes. The velocity of a phonon is in general a function of its subband, its frequency, and its *q*-point. The quantity in Eq. 11 effectively suppresses the subband and *q*-point indices and provides a quantity that is only frequency (or energy) dependent. Similar ""effective" quantities are also used in actual thermal conductivity calculations with adequate accuracy in the results [8, 28], although in our actual calculations we utilize all the information of the phonon spectrum.

In contrast to the sound velocity, the average weighted group velocity over different modes can provide a better and more correct insight with regards to phonon transport. Figure 7 shows that the average group velocity of phonons for NWs in the $\langle 110 \rangle$ is the highest, or high enough compared to the other orientations, in the largest part of the energy spectrum. The explanation behind this lies in the shape of the dispersions shown in Fig. 1. In the case of the <111> NW, the modes look overall less dispersive, i.e. more "flat". This results in the lowest average group velocity. The modes of the <110> NW, on the other hand, are more dispersive, i.e. less "flat", which results in the highest group velocity. By considering that all NWs with the same cross sectional area have the same phonon DOS, the differences in the group velocity explain the orientation dependence observed in the transmission functions and in the thermal conductance shown in Fig. 2 and Fig. 3, respectively. Eventually the transmission function of the <111> NW is ~1.25x lower than that of the <100> NWs and almost 2x lower than that of the <110> NWs.

## IV. Conclusions

In this work we study the thermal properties of ultra-narrow silicon nanowires using the atomistic modified valence force field method for the computation of the phonon bandstructure. We extract the thermal properties using the Landauer formalism under ballistic, and diffusive boundary-scattering-limited transport considerations. We address the effects of structural confinement on the phonon dispersion, the phononic density of states, the phononic transmission function, the sound velocity and the effective group velocity. Our results show that differently oriented NWs can have up to a factor of ~2x difference in their effective group velocity, transmission function and finally ballistic and diffusive thermal conductance. The <110> oriented NW has the highest ballistic and diffusive thermal conductance, followed by the <100> and finally the <111> NW. By taking into account the thermoelectric power factor, we show that the <111> orientation is the optimal choice for NW TE channels, since it provides the lowest thermal conductance and highest power factors for both *n*-type and especially *p*-type channels.

## Acknowledgements

This work was supported by the European Community's Seventh Framework Programme under grant agreement no. FP7-263306 (NanoHiTEC).

# References

[1] A. I. Boukai, Y. Bunimovich, J. Tahir-Kheli, J.-K. Yu, W. A. Goddard, and J. R. Heath, Nature, vol. 451, pp. 168-171, 2008.

[2] A. I. Hochbaum, R. Chen, R. D. Delgado, W. Liang, E. C. Garnett, M. Najarian, A. Majumdar, and P. Yang, Nature, vol. 451, pp. 163-168, 2008.

[3] I. Ponomareva, D. Srivastava, and M. Menon, Nano Letters, vol. 7, pp. 1155-1159, 2007.

[4] N. Yang, G. Zhang, and B. Li, Nano Letters, vol. 8, pp. 276-280, 2008.

[5] S.-C. Wang, X.-G. Liang, X.-H. Xu, and T. Ohara, J. Appl. Phys., vol. 105, p. 014316, 2009.

[6] M. Liangraksa and I. K. Puri, J. Appl. Phys., vol. 109, p. 113501, 2011.

[7] J. H. Oh, M. Shin, and M.-G. Jang, J. Appl. Phys., vol. 111, p. 044304, 2012.

[8] N. Mingo, Phys. Rev. B, vol. 68, p. 113308, 2003.

[9] X. Lu and J. Chu, J. Appl. Phys., vol. 100, p. 014305, 2006.

[10] M.-J. Huang, W.-Y. Chong, and T.-M. Chang, J. Appl. Phys., vol. 99, p. 114318, 2006.

[11] P. Martin, Z. Aksamija, E. Pop, and U. Ravaiolo, Phys. Rev. Lett., vol. 102, p. 125503, 2009.

[12] A. Paul, M. Luisier, and G. Klimeck, J. Appl. Phys., vol. 110, p. 114309, 2011.

[13] Z. Sui and I. P. Herman, Phys. Rev. B, vol. 48, pp. 17938-17953, 1993.

[14] A. Paul, M. Luisier, and G. Klimeck, J. Comput. Electron., vol. 9, pp. 160-172, 2010.

[15] H. Karamitaheri, N. Neophytou, M. Pourfath, and H. Kosina, J. Comput. Electron., vol. 11, pp. 14-21, 2012.

[16] S. Datta, Quantum transport: Atom to transistor, Cambridge University Press, 2005.

[17] R. Landauer, IBM J. Res. Dev., vol. 1, pp. 223-231, 1957.

[18] Z. Aksamija and I. knezevic, Phys. Rev. B, vol. 82, p. 045319, 2010.

[19] E. B. Ramayya, D. Vasileska, S. M. Goodnick, and I. Knezevic, J. Comput. Electron., vol. 7, pp. 319-323, 2008.

[20] J. M. Ziman, Electrons and Phonons: The Theory of Transport Phenomena in Solids (Clarendon, Oxford, 1962).

[21] X. Lu, J. Appl. Phys., vol. 104, p. 054314, 2008.

[22] A. J. H. McGaughey, E. S. Landry, D. P. Sellan, and C. H. Amon, Appl. Phys. Lett., vol. 99, p. 131904, 2011.

[23] M. Luisier, Phys. Rev. B, vol.86, p. 245407, 2012.

[24] N. Neophytou and H. Kosina, Phys. Rev. B, vol. 83, p. 245305, 2011.

[25] Neophytos Neophytou and Hans Kosina, Nano Lett., vol. 10, no. 12, pp. 4913-4919, 2010.

[26] T. Thonhauser and G. D. Mahan, Phys. Rev. B, vol. 69, p. 075213, 2004.

[27] X. Lu, J. H. Chu,, and W. Z. Shen, J. Appl. Phys., vol. 93, pp. 1219-1229, 2003.

[28] J. Zou and A. Balandin, J. Appl. Phys., vol. 89, p. 2932, 2001.

Figure 1:

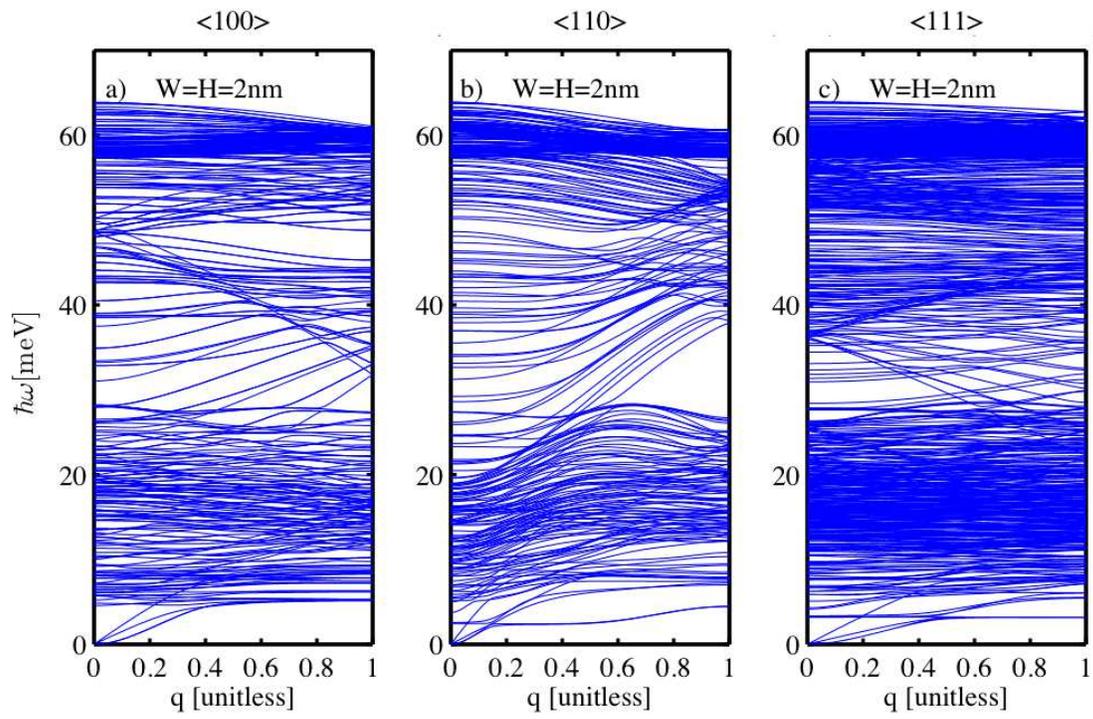

# Figure 1 caption:

Phononic dispersions of Si NWs of square cross sections with *W=H=2*nm for a) <100>, b) <110>, and c) <111> transport orientations.

Figure 2:

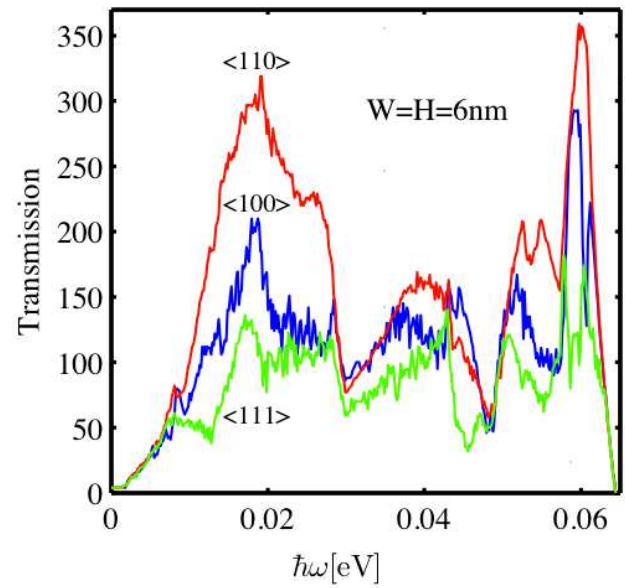

Figure 2 caption:

Transmission function versus energy for NWs in different transport orientations. <110> NWs have the highest and <111> NWs have the lowest transmission.

Figure 3:

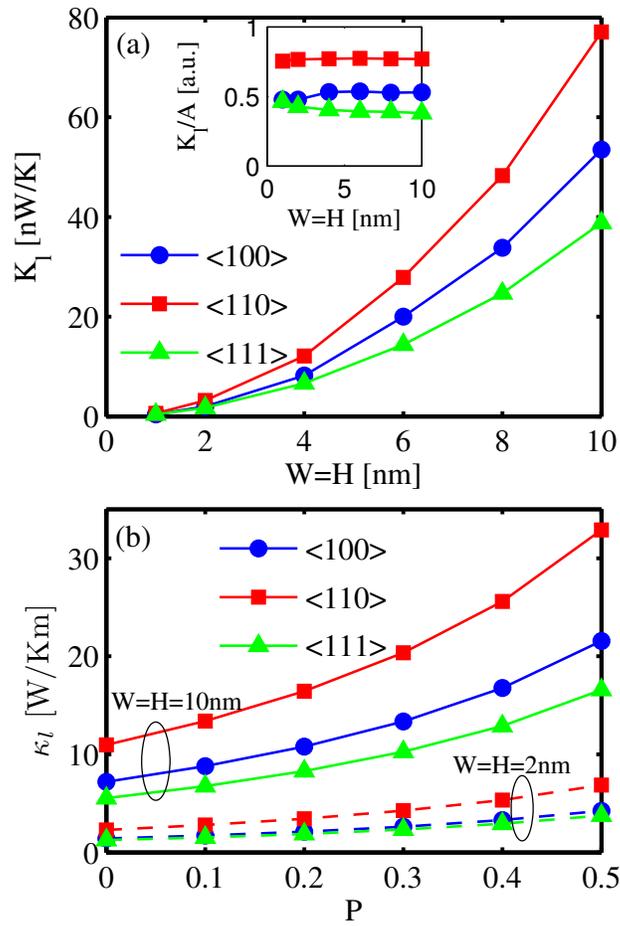

## Figure 3 caption:

(a) Ballistic lattice thermal conductance versus NW cross section side size for NWs in different orientations. Inset: The thermal conductance normalized by the cross section area. (b) Diffusive boundary-scattering-limited lattice thermal conductivity for NWs with different orientations versus the specularity parameter. Nanowires with side sizes of 2nm and 10 nm are considered.

Figure 4:

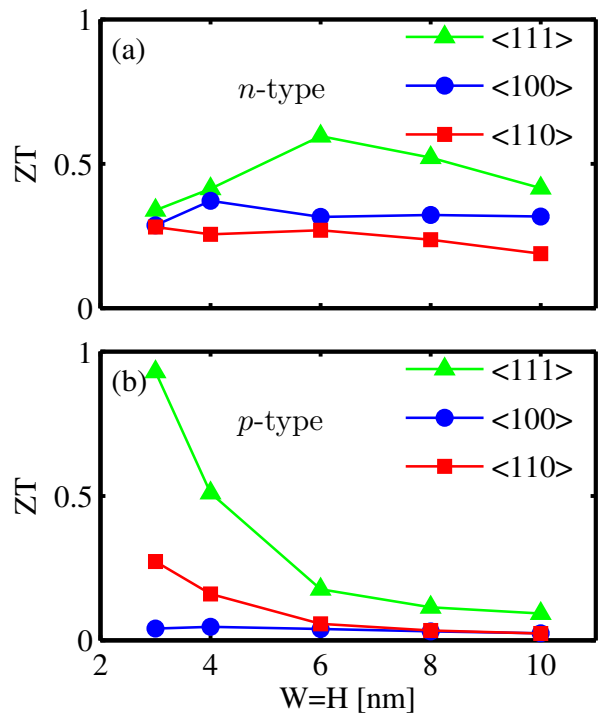

## Figure 4 caption:

The thermoelectric figure of merit *ZT* for (a) *n*-type and (b) *p*-type NWs with different orientations versus the nanowire side size. For this calculations, we used the thermoelectric power factors from Ref. [24].

Figure 5:

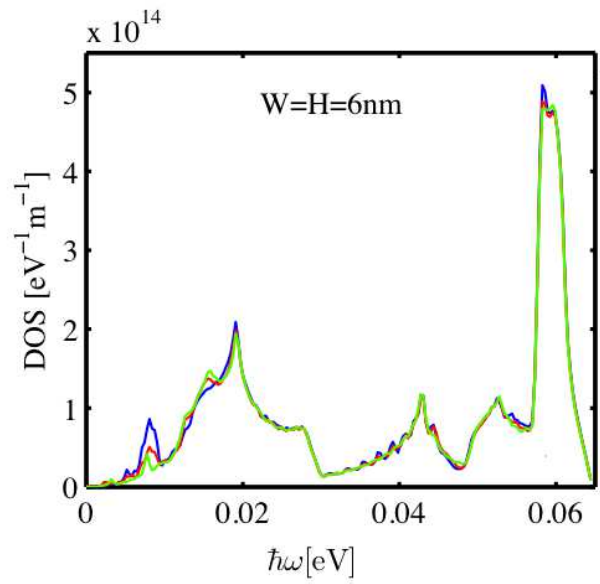

Figure 5 caption:

The phonon DOS versus frequency for NWs in different orientations. The DOS is nearly the same for the different transport orientations.

Figure 6:

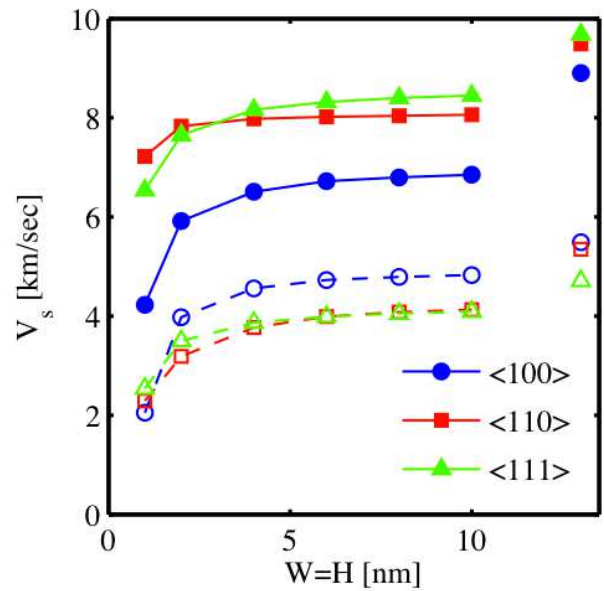

Figure 6 caption:

The sound velocity of NWs in different orientations versus the NW side size. Longitudinal and transverse velocities are shown in solid and dashed lines, respectively. Dots: The respective bulk Si velocities.

Figure 7:

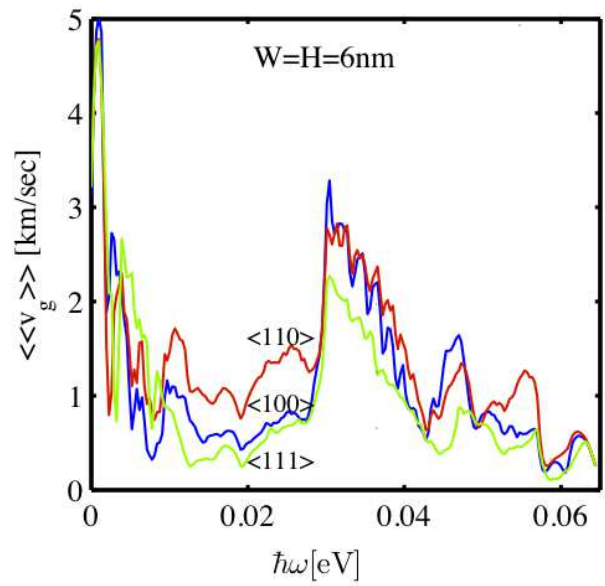

Figure 7 caption:

Average group velocity of NWs in different transport orientations versus frequency.